# Unveiling the Dynamic Infrared Sky with Gattini-IR

Anna M. Moore[1], Mansi M. Kasliwal[2], Christopher R. Gelino[3], Jacob E. Jencson[2], Mike I. Jones[4], J. Davy Kirkpatrick[3], Ryan M. Lau[5], Eran Ofek[6], Yuri Petrunin[7], Roger Smith[1], Valery Terebizh[8], Eric Steinbring[9] and Lin Yan[3]

[1]*Caltech Optical Observatories, California Institute of Technology, 1200 E California Blvd., Mail Code 11-17, Pasadena, CA 91125*
[2]*Division of Physics, Math and Astronomy, California Institute of Technology, 1200 E California Blvd., Mail Code 249-17, Pasadena, CA 91125*
[3]*Infrared Processing and Analysis Center, California Institute of Technology, 770 S. Wilson Avenue, MS 100-22, Pasadena, CA 91125*
[4]*Precision Optics of Azle, LLC, 816 Wayne Trail, Azle TX 76020*
[5]*Jet Propulsion Laboratory, California Institute of Technology, 4800 Oak Grove Drive, Pasadena, CA 91109*
[6]*Weizmann Institute of Science, Rehovot 76100, Israel*
[7]*Telescope Engineering Company, 15730 W. 6-th Ave., Golden, CO 80401*
[8]*Crimean Astrophysical Observatory, Nauchny, Crimea, Ukraine*
[9]*National Research Council Canada, Herzberg Astronomy and Astrophysics, 5071 West Saanich Road, Victoria, British Columbia, Canada V9E 2E7*

## ABSTRACT

While optical and radio transient surveys have enjoyed a renaissance over the past decade, the dynamic infrared sky remains virtually unexplored. The infrared is a powerful tool for probing transient events in dusty regions that have high optical extinction, and for detecting the coolest of stars that are bright only at these wavelengths. The fundamental roadblocks in studying the infrared time-domain have been the overwhelmingly bright sky background (250 times brighter than optical) and the narrow field-of-view of infrared cameras (largest is 0.6 sq deg). To begin to address these challenges and open a new observational window in the infrared, we present Palomar Gattini-IR: a 25 sq degree, 300mm aperture, infrared telescope at Palomar Observatory that surveys the entire accessible sky (20,000 sq deg) to a depth of 16.4 AB mag (J band, 1.25um) every night. Palomar Gattini-IR is wider in area than every existing infrared camera by more than a factor of 40 and is able to survey large areas of sky multiple times. We anticipate the potential for otherwise infeasible discoveries, including, for example, the elusive electromagnetic counterparts to gravitational wave detections. With dedicated hardware in hand, and a F/1.44 telescope available commercially and cost-effectively, Palomar Gattini-IR will be on-sky in early 2017 and will survey the entire accessible sky every night for two years. We present an overview of the pathfinder Palomar Gattini-IR project, including the ambitious goal of sub-pixel imaging and ramifications of this goal on the opto-mechanical design and data reduction software.

Palomar Gattini-IR will pave the way for a dual hemisphere, infrared-optimized, ultra-wide field high cadence machine called Turbo Gattini-IR. To take advantage of the low sky background at 2.5 um, two identical systems will be located at the polar sites of the South Pole, Antarctica and near Eureka on Ellesmere Island, Canada. Turbo Gattini-IR will survey 15,000 sq. degrees to a depth of 20AB, the same depth of the VISTA VHS survey, every 2 hours with a survey efficiency of 97%.

*amoore@astro.caltech.edu; phone 1 626 395-8918; fax 1 626 568-1517;

# 1. INTRODUCTION

Exploring the transient sky at infrared wavelengths, the Palomar Gattini-IR ultra-wide infrared telescope will be located at the historic Palomar Observatory (Figure 2a). The telescope has an aperture size of 300mm and a capture field of 25 square degrees. The system operates in the J band, at a wavelength of 1.25 μm, where the near-infrared sky is darkest from this temperate location. The system design is based on the immediate availability of two critical in-house components: an engineering grade infrared detector from Teledyne, called a Hawaii 2RG, with 2048 by 2048 active pixels (Figure 2f), and a ND5 cryogenic dewar from Infrared Laboratories (Figure 2e).

Gattini-IR will survey the northern night sky to a depth of 16.4AB every night, with a survey time efficiency of 94%. This survey will repeat continuously under all observable sky conditions, each new epoch adding to an increase in survey sensitivity until the extragalactic confusion limit is reached within 24 months (20.1AB). The baseline survey will operate for 24 months, in addition, to ensure parallel observations with two successive Advanced LIGO runs. Palomar Gattini-IR will probe the infrared signal of every visible object in the hemisphere on time scales of hours, days, weeks and months. It will capture rare, infrared events in the local Universe. Finally, operating synergistically with the upcoming Zwicky Transient Factory (May 2017 onwards), the combination of the optical and infrared data will produce multi-color time series data with an unprecedented wealth of information vital for object classification.

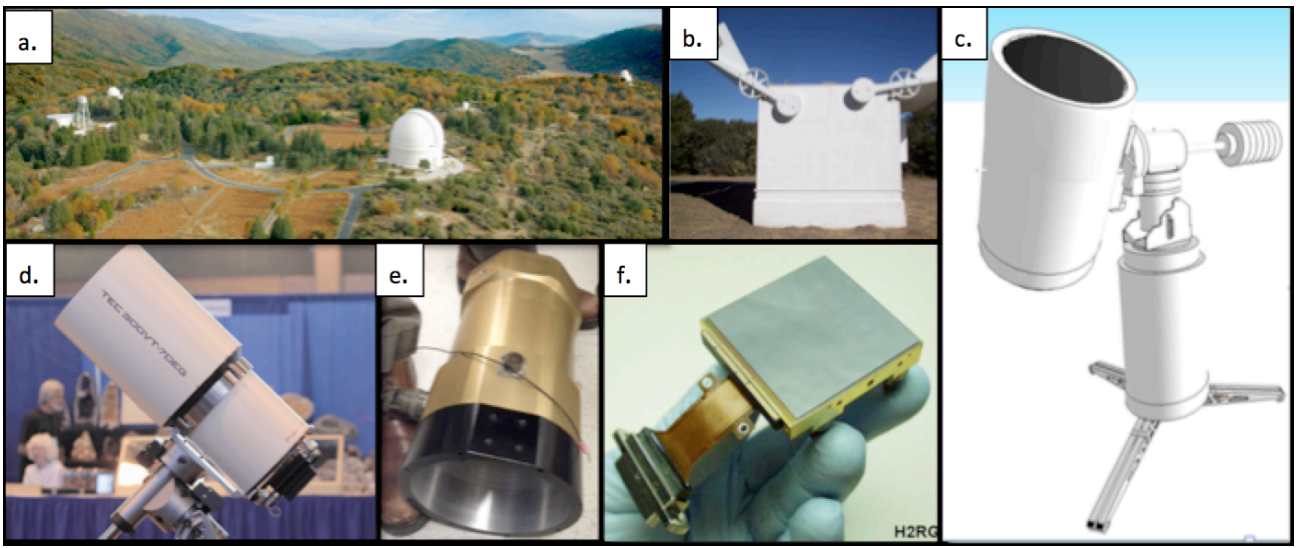

Figure 1: (a.) An aerial view of Palomar Observatory, Temecula, CA. (b.) The instrument will be housed in an existing dome at the Observatory. (c.) A rendering of the Gattini-IR ultra-wide field infrared telescope. (d.) The F/1.44 TEC-300VT telescope. (e.) The in-house ND-5 large cryogenic dewar from Infrared Laboratories. (f.) The in-house 2kx2k Hawaii 2RG infrared detector from Teledyne.

# 2. SCIENCE CASE

While optical wavelengths are a powerful band with which to explore thermal transients, and radio bands are well-suited for sources with relativistic ejecta, both are blind or insensitive to transients that are either self-obscured or located in dusty regions (e.g., molecular clouds, dense starburst regions). We are now aware of *at least* three new classes of explosive transients where the bulk of the emission is in the infrared [1]. We also know that a large fraction of luminous stars are self-obscured as they are intrinsically unstable due to radiation pressure and/or copious mass-loss and circumstellar dust formation. Moreover, recent opacity calculations [2] suggest that the spectra of electromagnetic counterparts to gravitational wave sources such as neutron star mergers peak in the infrared.

Here, we present some science highlights expected from Palomar Gattini-IR, in order of low-risk to high-risk.

## 2.1 Supernovae and classical novae

Palomar Gattini-IR will discover hundreds of supernovae and tens of classical novae every year. It will be sensitive to all transients, even those that are heavily enshrouded with more than 10mag of extinction and completely missed by optical surveys. This science case is the "bread and butter" science for this survey.

## 2.2 Monitoring of cool stars

Palomar Gattini-IR will detect 366 Brown Dwarfs in every scan of the sky, a total of 400 times for every Brown Dwarf, over the two-year life of the survey. Faster cadences and higher sensitivities are possible using targeted observing programs. Palomar Gattini-IR is well matched to this science case as Brown Dwarfs, while optically very faint, are extremely bright in the infrared. We predict major advances in the area of statistical analysis of weather on such stars, detection of planetary transits and possibly even discovery of new Brown Dwarfs using the parallax method. Transits by gas giant planets will be easy to detect since the brown dwarf and the giant planet have similar sizes.

## 2.3 Stellar mergers

Gattini-IR will uncover stellar mergers, at the low mass end in our own Galaxy and at the high-mass end in other galaxies. Coalescence of 1—30 $M_\odot$ binaries is expected to create large amounts of dust in an optically thick wind launched during the stellar merger [3].

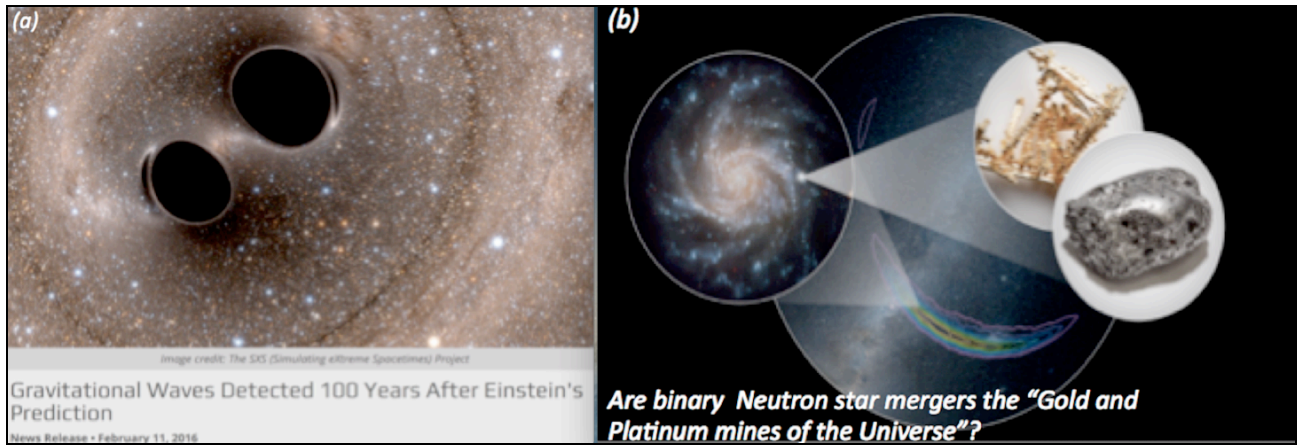

Figure 2: (a) The detection of gravitational waves by Advanced LIGO was announced on Feb 11, 2016. (b) Gattini-IR would search gravitational wave error circles to look for the formation sites of gold, platinum and other heavy elements in the Universe.

## 2.4 Infrared signatures of gravitational wave events

Since gravitational wave events are coarsely localized to hundreds of square degrees [4], and, given the recent spectacular announcement of the first gravitational wave detection from the Advanced LIGO consortium [5], we urgently need a wide-field infrared telescope that will promptly slew to look for an electromagnetic counterpart. Gattini-IR could pinpoint the infrared counterpart of a neutron star merger if the source is brighter than 20AB (J band; 5 sigma), assuming a median localization of 250 square degrees and a 5 night exposure. Transient detections from Gattini-IR will be followed up by spectroscopic facilities such as the Palomar 200-inch and Keck. These data will begin to test the theory that such mergers are the production sites in the Universe for rapid-neutron capture elements such as gold and platinum (see Figure 1) [6].

## 2.5 Birth of black holes

Gattini-IR would be sensitive to massive stars collapsing directly to form black holes. Theoretically, weak shocks in failed supernovae that form black holes may also lead to bright infrared, rather than optical, transients; the ejection of large amounts of material at low velocity may condense to form dust. No observational evidence has been seen thus far of such an event, due in part to the present lack of rapid-scan, wide-field J-band instrumentation as proposed herein.

.

## 3. SCIENCE FLOWDOWN

The science case is implemented by a wide field, infrared, telescope of ~300mm aperture. The aperture is set by the need of several science cases to probe point source sensitivities of $16M_{AB}$ over wide areas per night (e.g. cool stars). This performance results in a point source sensitivity approaching $20M_{AB}$ for a 5 night integration over 250 square degrees, a minimum required for detection of, for example, electromagnetic counterparts of neutron star mergers. The field of view is, to a large extent, set by technology. The dedicated single Hawaii-2RG infrared array is fed by a commercial F/1.44 telescope, resulting in a field of view of 25 sq. deg, with a pixel size on sky of ~8.6 arcsec. The most stringent requirement of the instrument design and data analysis is the requirement for accurate photometry with sub-pixel imaging to increase the instrument point source sensitivity in this highly background limited application. A widening of the image point spread function to the Nyquist sampling of 2x2 pixels results in a fourfold increase in the noise contribution, or just under a magnitude decrease in the sensitivity of the nightly survey. In addition, the effect of source confusion is doubled in severity in this already large plate scale regime. We discuss the implications of the requirement for sub-pixel imaging in Section 6.2. Table 1 presents the top level specifications of Palomar Gattini-IR and corresponding point source sensitivity assuming sub-pixel imaging performance.

Table 1. Palomar Gattini-IR summary and specifications.

| **Palomar Gattini-IR** | |
|---|---|
| Telescope Aperture | 300mm |
| Telescope F/ratio | 1.44 |
| Field of view | 25 sq. deg |
| Detector | Teledyne Hawaii 2RG |
| Pixel size/format | 18 μm/2048x2048 |
| Plate scale/pixel | 8.59 arcsec |
| Filter | 2MASS J band |
| Limiting magnitude (SNR=5, t=30s, sky=$15.6M_{Vega}$arcsec$^{-2}$) | $16.4M_{AB}$ |
| Limiting magnitude (SNR=5, t=10,000s sky=$15.6M_{Vega}$arcsec$^{-2}$) | $19.5M_{AB}$ |

## 4. OPTICAL DESIGN

The first approach to produce a viable optical design for Palomar Gattini-IR was a custom refractive three element design based on a Mangin system. The resulting 300mm F/1 layout required one aspheric surface on the front element and two aspheric surfaces on the field flattener element located close to the detector to enable the sub-pixel imaging over, in this case, 50 square degrees. In addition, the layout located the infrared detector within the telescope barrel, which increased the complexity of the cooling system needed to keep the infrared array cryogenically cold in this otherwise non-cryogenic system.

A more cost-effective baseline solution was found in a commercial telescope from Telescope Engineering Inc. called the Terebizh TEC300VT. This impressive 300mm aperture F/1.44 system contains 6 all-spherical elements with sub-pixel imaging in the visible spectrum across the whole field. The optical design required only slight re-spacing to achieve sub-pixel image quality over the entire format and J-band spectrum, and over a 0-30ºC temperature range with refocusing. The modified optical design is shown in the left-hand image of Figure 3. It contains five stray-light baffles located in front of the entrance pupil, six optical elements, a vacuum window and a J-band cold filter and detector at cryogenic temperature.

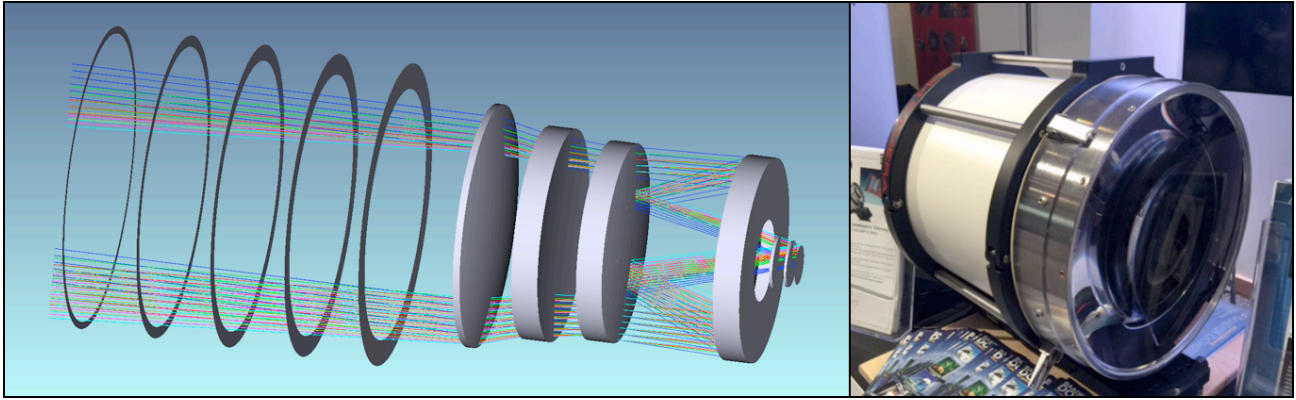

Figure 3: [Left] The TEC300VT contains six spherical elements and is shown with the additional dewar window and filter, as well as five baffles located in front of the telescope pupil. [Right] The fully assembled TEC300VT prototype telescope.

### 4.1 Optical layout

The TEC300VT is optimized for visible imaging and required minor alteration in the element spacing to optimize the performance for this J band application. The optical prescription of this off the shelf system is proprietary, however, the lens element schematic for the J band optimized TEC300VT is shown in Figure 4. The telescope is an all spherical system. The larger elements are made of N-BK7, while the doublet is a combination of N-BAF4 and N-LASF46A. The lens elements will be coated with a standard NIR coating optimized for the J band.

### 4.2 Image quality

The imaging performance of the optimized TEC300VT, combined with the dewar window and filter, is shown in the spot diagrams of Figure 5. The spot diagrams are shown as a function of field position and ambient temperature and are excellent. Detector refocus is implemented during temperature changes. The point spread function is very uniform across the field and as a function of temperature.

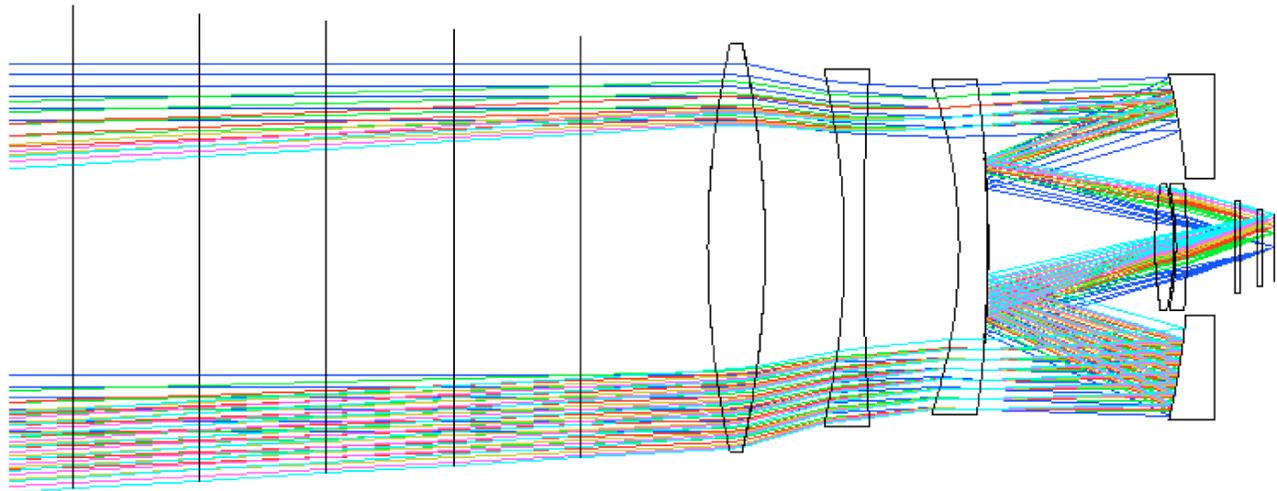

Figure 4: Lens drawing with rays incorporated of the TEC300VT Palomar Gattini-IR system with the front 5 light baffles shown.

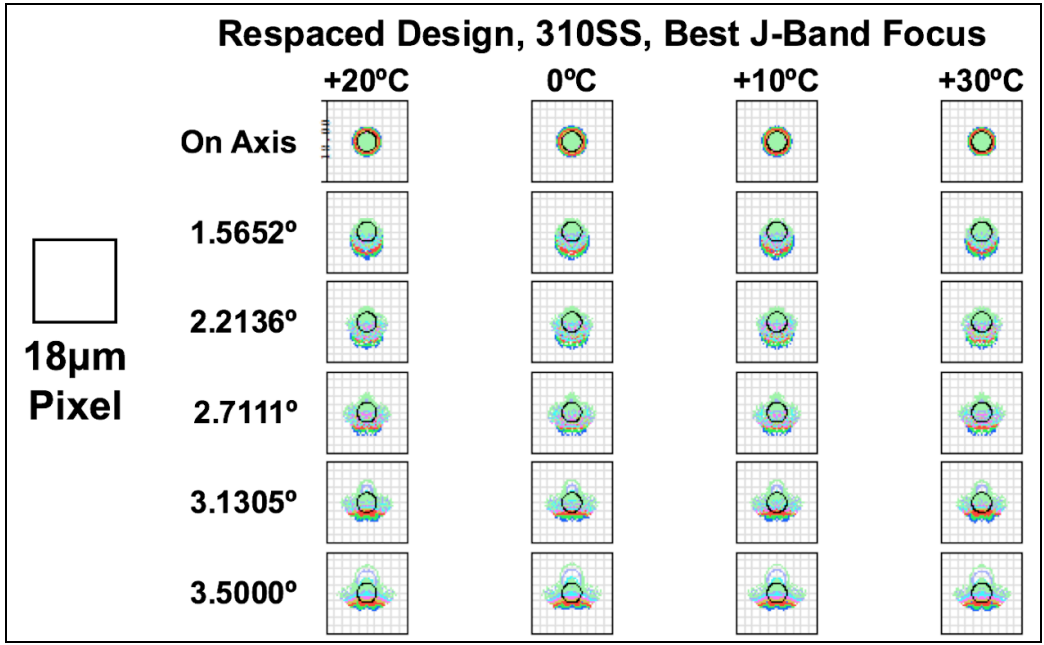

Figure 5: Spot diagrams across the field for the optical prescription, shown relative to the detector pixel size and as a function of temperature assuming detector refocus. The Airy circle for this system is represented by a black circle.

### 4.3 System throughput

The system throughput allocated to the atmosphere, telescope, dewar optics and detector is ~26%, assuming 0.25% loss per air-glass interface and mirror reflection, 30% loss at the detector and in-band filter transmission of 95%. The telescope itself suffers additional vignetting towards the edge of the field of view, with a maximum loss of ~10% at the corners of the 5º x 5º format.

### 4.4 Stray light and multi-bounce ghost analysis

The sky emission at 1.25μm, at ~15-16$M_{Vega}$arcsec$^2$, is 4-5 times brighter than at visible wavelengths. A preliminary stray light and ghost analysis of the modified TEC300VT design at the focal plane was performed using Lambertian background scattering to evaluate the magnitude of ghost intensities surrounding an on-axis image using a non-sequential Zemax [8] optical model. The model investigated three coating scenarios for the lens elements: uncoated optics, single-layer AR coating and a more costly high efficiency multi-layer AR coating. Figure 6 gives a summary of the results. The lower three plots in this figure show the intensity distribution surrounding the on-axis image for the three coating cases. To aid in the comparison, intensity cross sections of the three cases are shown in the upper plot of Figure 6. The single layer AR coating produces a ghost intensity of $10^{-6}$ compared to the on-axis image, which is perfectly adequate for this application. Furthermore, the results support the use of the manufacturer-provided baffles located ahead of the entrance pupil, as well as proper application of subaperture, ultra-flat black, highly J-band absorbent lens surface coatings, for further reducing stray light contribution. Analysis is on-going to investigate off-axis scattering properties, both within and outside of the field.

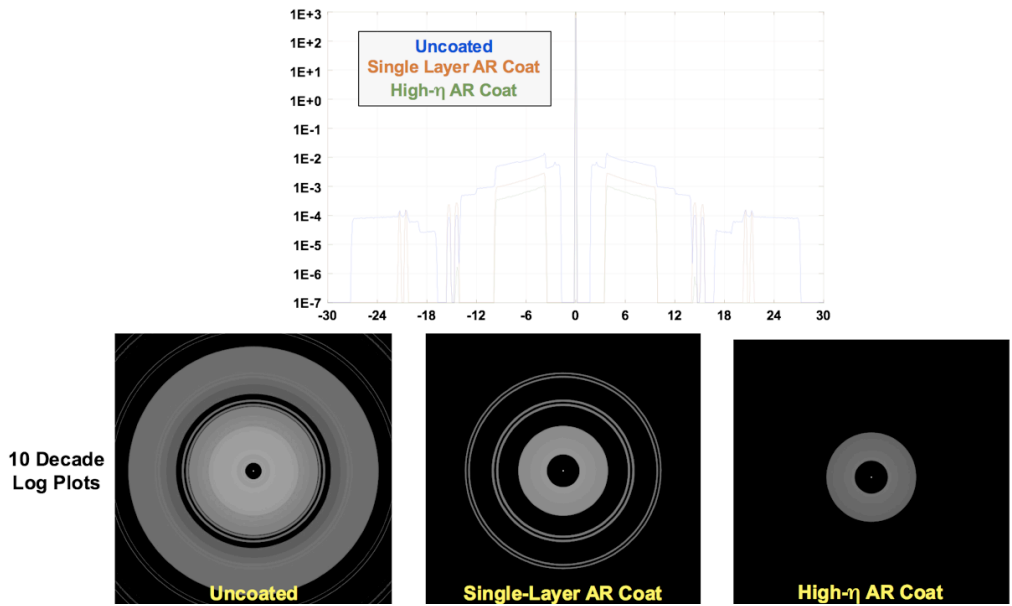

Figure 6: Stray-light analysis results for three optical coating cases. Single layer AR coatings produce ghost images that are $10^{-6}$ less than the on-axis image, which is an adequate result for Palomar Gattini-IR.

### 4.5 Temperature and focus

As the modified TEC300VT optical elements and opto-mechanical structures thermally equilibrate to a given ambient operational temperature at the Palomar site, the position of best f/1.44 focus changes and must be actively compensated for. The focus airspace variation linearly decreases with increasing soak temperature, as shown in Figure 7. Focus shift amounts to about -110μm as the soak temperature rises from 0ºC to 30ºC. This analysis shows the need for active focus compensation of the detector dewar assembly relative to a fixed telescope barrel to the level of ±10μm. Transient thermal changes are computationally difficult and were not modeled here, but the small telescope size and metal barrel construction should allow for rapid thermal equilibration. A more detailed thermal analysis is underway by creating a finite element model of the combined telescope and detector housing assembly.

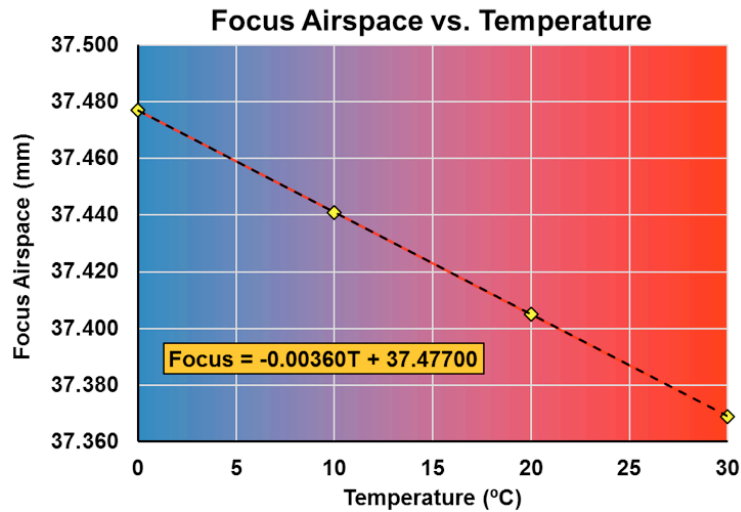

Figure 7: Variation of the best focus location with thermally equilibrated telescope structure temperature.

# 5. MECHANICAL DESIGN

## 5.1 Detector housing

The detector will be housed in an existing ND2 liquid nitrogen dewar from Infrared Laboratories. The housing, shown in the left-hand image of Figure 8, requires the following additions and modifications: (i) a vertical fill tube will be inserted to permit operation on the dewar in a Cassegrain configuration on an equatorial mount; (ii) the mechanical mount for the Hawaii-2RG array requires design, thermal and structural analysis and fabrication and (iii) the mechanical structure and mechanism linking the dewar to the telescope barrel requires design, analysis and fabrication. The dewar will require refilling with liquid Nitrogen once a day through the life of the survey.

## 5.2 Telescope barrel

The mechanical design of the telescope barrel is shown in the right-hand image of Figure 8. The lens element groups are housed in titanium cells that are spaced using invar rods. The use of 301 stainless steel rods rather than invar is currently under investigation as the expansion of this material nicely compensates the temperature induced refractive index and dimensional changes of the lens groups. The assembled barrel and detector housing will be modeled and analyzed in Solidworks to confirm thermal and structural requirements placed on the system.

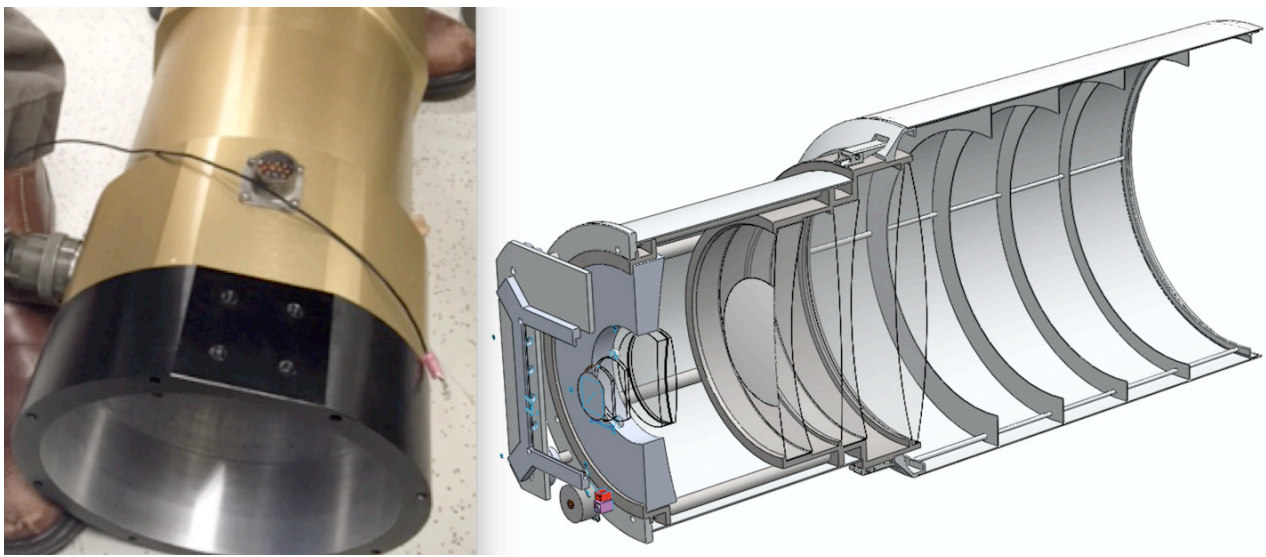

Figure 8: [left] An in-house ND2 liquid nitrogen dewar with custom extension will house the detector and filter. [right] The mechanical model of the TEC300VT is shown in sectional view including the front baffle.

## 5.3 Tracking mount

A commercial tracking mount from Software Bisque was chosen for this application. This sturdy mount, model Paramount MEII and shown on the left-hand side of Figure 9, can support up to 240lbs of instrument mass and an additional 240lbs of counterweight mass, sufficient for this application. The slewing speed is very important for this infrared application, as any additional time lost to field repointing directly results in time not on sky, given the readout rates for a single Hawaii-2RG are a few seconds only. The raw pointing of the mount across the whole sky after software modeling, equates to ½-1 pixel, with the additional option to implement encoders on each axis that would result in sub-pixel accuracy. A trade study is underway to investigate the implementation of axis encoders versus a slower, parallel, optical guider. The mount is provided with software to operate the mount, and such a system has recently been installed at Palomar Observatory for another telescope project.

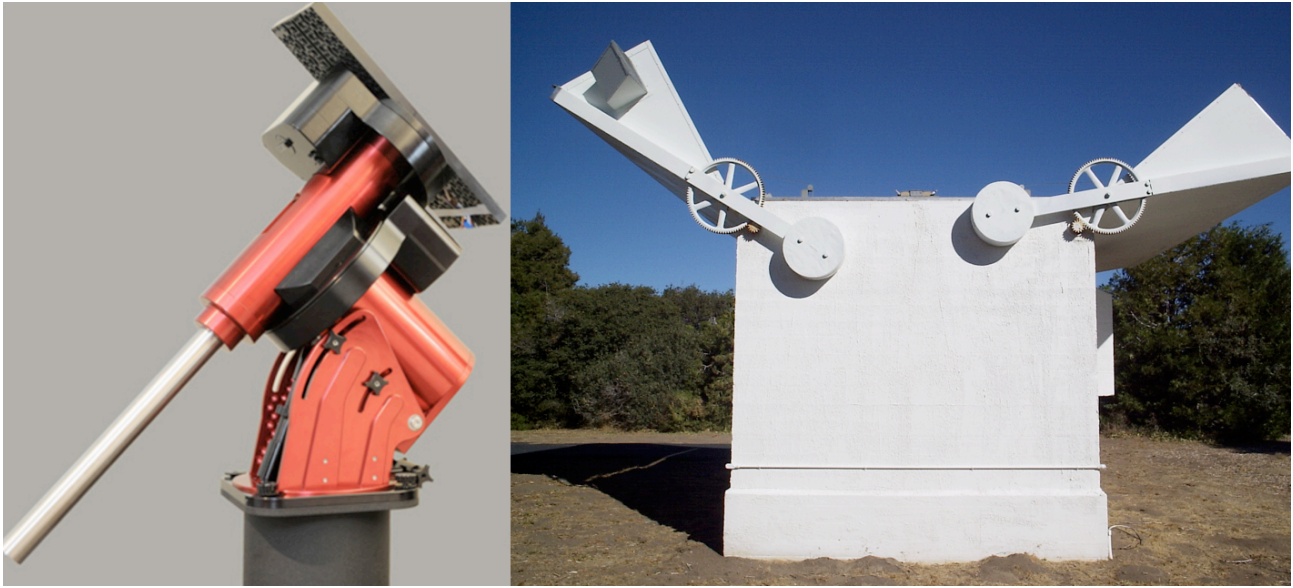

Figure 9: [Left] The Paramount II telescope tracking mount from Software Bisque. [Right] Palomar Gattini-IR will be located in an existing housing at Palomar Observatory.

### 5.4 Housing

Palomar Gattini-IR will be located in an existing housing at Palomar Observatory, shown in the right-hand image of Figure 9. The housing has an existing concrete pier for mounting, with access to power and network. The Observatory staff will provide the necessary Liquid Nitrogen fill approximately once per day.

## 6. SURVEY AND DATA REDUCTION

### 6.1 Baseline survey

The baseline survey will be a shallow, 20,000 square degree survey acquired every night. Each 25 square degree field is observed for a total exposure time of 40 seconds that is split between two 20 second exposure dithers to provide adequate sky subtraction. The temporal overhead allocated for repointing between fields is 2 seconds that is deliberately matched by the time required to read-out and store the individual reads. This nightly survey requires 800 reduced tiles, and requires a little over 7 hours to complete. The survey efficiency is 94%. Each nightly survey will achieve a depth of $16.4M_{AB}$ on point sources with a signal to noise ratio of 5. Stacked nightly surveys obtain deeper sensitivities, so cadence must be balanced against depth. For example, stacked weekly surveys can in principle obtain a depth of $17.5M_{AB}$, whereas stacked monthly surveys can obtain a depth of $18.3M_{AB}$. The extragalactic confusion limit is a little over $20M_{AB}$, marking the maximum depth for this instrument outside of the galactic plane.

Surveys of varying area, depth and cadence will be chosen in addition to the baseline survey that best match the science case.

### 6.2 Data reduction

The goal of achieving sub-pixel imaging places very tight requirements on the data reduction and instrument performance. Several methods are under investigation for how best to implement the data reduction and we only briefly introduce them here. On-site commissioning of the telescope mount using an under-sampled CCD prior to the arrival of infrared telescope can be used to select and mature the best observational and data reduction techniques.

(i) Use of Reference images

The primary approach is the develop a set of reference images that can be used to compare with on-sky survey subtracted dithers. For this approach to be successful, we must ensure as best as possible that the instrument performance is uniform across varying observing conditions such as pressure changes, zenith distance or temperature changes. We have begun to document such conditions and to model their effects on the PSF shape or location, and they are listed versus severity in Figure 10. It is possible that uniform Gaussian images can be repeatedly produced using the telescope drives during observation, whilst still obtaining sub-pixel imaging.

(i) Drizzling

In the case that the reference images are not successful, there is the option of using the telescope drives to deliberately smear the images during each 15 second exposure across two pixels. The same approach can be used during the second dither, in an orthogonal direction to the first. This approach is a factor of sqrt(2) less sensitive that the sub-pixel imaging application, however, is it more sensitive than increasing the image to fill the nominal 2x2 pixels, or Nyquist sampling.

# 7. TURBO GATTINI-IR

Palomar Gattini-IR is a pathfinder for a more ambitious project for the near future called TURBO Gattini-IR. This instrument takes advantage of the colder atmosphere at a polar location such as Ellesmere Island in the High Arctic, or the South Pole station in Antarctica, where the atmospheric thermal emission is shifted red-ward enough to reveal a very low sky background at 2.4µm. The instrument consists of four identical 0.5 m aperture barrels, is shown in Figure 11. A comparison of the survey capability to that of the VISTA Ks survey is given in Table 2 . The instrument is capable of surveying the entire sky to a depth equivalent to the VISTA VHS survey at $K_S$, in 2 hours. In order to achieve this performance, an equivalent of forty 4k x 4k infrared arrays, or equivalent, are required with a pixel sampling of 1.38 arcsec. The field of view of one image is 100 square degrees. Spectacular advances from both the transient and static surveys are predicted with this instrument, that offers strong synergies with existing and future facilities such as LSST, JWST and WFIRST.

| PSF modeling across the full field of Palomar Gattini-IR | | | |
|---|---|---|---|
| Column | Term | Description | Method of analysis |
| 1 | Optical aberrations over FOV, J-Band spectrum | Lens element manufacturing tolerances, assembly positional tolerances. | Contact vendor to get tolerance specifications, apply to nominal prescription with Sensitivity Analysis, Monte Carlo Gaussian statistical analysis. Obtain as-manufactured parameters for each element (radii, thickness, diameters, surface irregularity, surface scattering TIS values, glass melt data, residual wedge, AR coating transmission vs. incidence angle) |
| 2 | Temperature / Pressure variations | Changes in ambient temperature and atmospheric pressure increases or decreases air refractive index relative to lens element refractivity. Glass materials change refractivity over temperature (dn/dt). Glass elements shrink or grow according to their thermal coefficients of expansion. Tube assembly shrinks or grows according to material thermal coefficient of expansion. | Zemax accounts for variations in air refractivity with temperature and pressure. Contact vendor to get latest mechancal design and run a thermal FEA model to get impact on element positions. Incorporate results into zemax. |
| 3 | Focal plane irradiance over FOV | Large range of incidence angles over FOV on optical surfaces due to low system focal ratio. System is not telecentric, therefore cosine falloff is expected. Wavelength-dependent changes in system transmission over FOV. | Analyze in Zemax. Need AR coating properties over range of incidence angles, J-band spectrum. |
| 4 | Stray light analysis | Stray light due to direct or scattered radiation reduces image contrast over FOV. | Analyze in Zemax using Non-Sequential Surface analysis. Determine optimum placement of baffles, surface blackening. Explore Vantablack as candidate light absorbing material. |
| 5 | Gravitational sag | Possible changes in PSF patterns across FOV due to instrument deflection versus operational orientation at 1g. | Contact vendor to get latest ME design and create a dynamic FEA model with varying instrument orientation. Incorporate deflection analysis results into Zemax model. |
| 6 | Seeing | Blurring of image due to time-integrated atmospheric turbulence. | Incorporate Palomar seeing measurements in the visible, scale to J band and include in zemax model using FWHM |
| 7 | Atmospheric Distortion over FOV | Changes in distortion across instrument FOV in vertical direction due to varying atmospheric refractivity (function of temperature and altitude). | Apply Zemax model for atmospheric refractivity across system FOV at different zenith angles. |
| 8 | Earth's motion round the Sun | Stellar aberration due to relative motion of Earth and source, predictable | Predictable - TBD effects on LOS pointing, tracking error; distortion across FOV. |
| 9 | Atmospheric Dispersion | Atmospheric dispersion in vertical direction across J-Band as a function of zenith distance over FOV. | Can be directly modelled in Zemax. Effects analyzed, found to be minimal due to J-band being in the infrared, and due to the short EFL of this instrument. |

Figure 10: A list of terms that are input into the PSF modeling, given with a brief description, and method of analysis. Low numbered items correspond to the most severe terms.

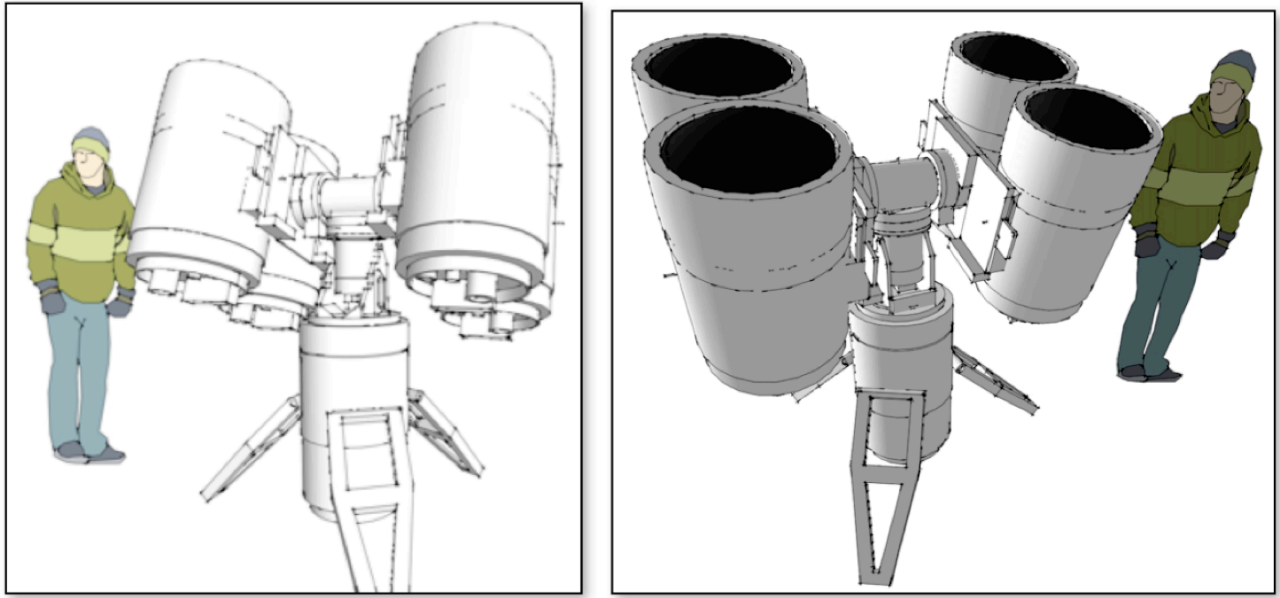

Figure 11: TURBO GATTINI-IR is a four barrel $K_{Dark}$ (2.4µm) imager, located at a cold site such as Ellesmere Island in the High Artic, or the South Pole station in Antarctica.

Table 2: Comparison of the VISTA VHS survey at $K_S$ to TURBO Gattini-IR at $K_{Dark}$.

|  | VISTA VHS (Ks) | TURBO GATTINI-IR ($K_{Dark}$) |
| --- | --- | --- |
| Baseline survey Area (square deg) | 18,000 | 15,000 |
| Baseline survey time to $20M_{AB}$ | ~600 hours[6] | 2 hours |
| Static survey depth (1 season) | $20M_{AB}$ | $23.4M_{AB}$ |

## 8. SUMMARY

We present a 25 square degree, J band imager called Palomar Gattini-IR, that will survey the observable sky to a depth of $16.4M_{AB}$ every night over a 2-year survey duration. The science case enabled by this pathfinder includes detection of the missing novae and SNe in optically opaque areas such as the galactic plane, monitoring of hundreds of northern hemisphere brown dwarfs, and potentially locating the electromagnetic counterparts of gravitational wave detections. The instrument is in fabrication and will begin commissioning at Palomar Observatory in the first quarter of 2017. This is the first step towards a more capable polar instrument, exploiting KDark and achieving 100 square degree FoV. That instrument is called Turbo Gattini-IR that will survey the northern and southern polar skies to a depth of $20M_{AB}$ every 2 hours taking advantage of the cold sky emission at these wavelengths. Both instruments aim to show the science enabled by small, wide field infrared cameras that are extremely agile in field acquisition and pointing.

## AKNOWLEDGEMENTS

We acknowledge and greatly thank the Mt. Cuba Astronomical Foundation and the California Institute of Technology for the generous support of this project. We greatly thank Palomar Observatory for the opportunity to install Palomar Gattini-IR at this historic site and the donation of in-kind hardware.